\documentclass[aps,prl,reprint,superscriptaddress,nofootinbib]{revtex4-2}

\usepackage{amsmath,amssymb,mathtools,bm}
\usepackage{booktabs}
\usepackage{microtype}
\usepackage{xcolor}
\definecolor{cbBlue}{HTML}{0072B2}\definecolor{cbOrange}{HTML}{E69F00}
\definecolor{cbGreen}{HTML}{009E73}\definecolor{cbVerm}{HTML}{D55E00}
\definecolor{cbPurple}{HTML}{CC79A7}\definecolor{cbSky}{HTML}{56B4E9}
\usepackage{graphicx}
\usepackage{tikz}
\usepackage{pgfplots}
\usetikzlibrary{decorations.pathmorphing,arrows.meta,calc,patterns}
\pgfplotsset{compat=1.18}
\usepackage[colorlinks=true,allcolors=blue!55!black]{hyperref}
\hypersetup{pdftitle={Hartree Fragmentation and Long-Range Pair Networks in Aperiodic Chains},
pdfauthor={Yogeshwar Prasad},
pdfsubject={Many-body localization; slowly varying potentials; long-range interactions},
pdfkeywords={many-body localization, slowly varying potential, long-range interactions, resonances}}

\newcommand{\dd}{\mathrm{d}}
\newcommand{\ii}{\mathrm{i}}
\newcommand{\order}{\mathcal{O}}

\begin{document}

\title{Hartree Fragmentation and Long-Range Pair Networks in Aperiodic Chains}

\author{Yogeshwar Prasad}
\email[Contact author: ]{yogeshwar@snu.ac.kr}
\affiliation{Department of Physics, Hanyang University, Seoul 04763, Korea}
\affiliation{Research Institute of Basic Sciences, Seoul National University, Seoul 08826, Korea}
\date{\today}

\begin{abstract}
For fermions in a slowly varying aperiodic potential with random
interactions $V_{ij}/|i-j|^\alpha$, the potential fixes the resonance supply
$N_0\propto L^{2-n}/h$.  On crossing $\alpha=2n$, wherever Hartree fragmentation is operative, the
resonant-object ensemble reconstructs: turning-point runs fragment into clusters whose intrinsic
matching law resolves no logarithmic singularity, while isolated wing bonds survive at finite
density and restore a reduced logarithm.  The scaling
$x_{\rm pair}^{4-2n-\alpha}\ln x_{\rm pair}\propto h^2/(Vt_0)$ is unchanged.  Random coefficient
fluctuations sustain the $R^{-\alpha}$ coupling; uniform ones cancel to $R^{-\alpha-2}$.
\end{abstract}

\maketitle

\emph{Introduction.}---Whether localization can survive power-law interactions is a
long-standing question~\cite{Burin2006,Yao2014,Burin2015,Hauke2015,NandkishoreSondhi2017,DeTomasi2019,Sierant2019,Thomson2020,Maksymov2020,Botzung2021,Vu2022,Defenu2023}.  Resonance-counting
arguments predict an instability of many-body localization (MBL)~\cite{Basko,Abanin_rev,Sierant_rev} for interactions decaying slower than $R^{-2d}$ in
$d$ dimensions~\cite{Burin2006,Yao2014}, but they assume independent local degrees of freedom with
spatially homogeneous resonance statistics.  Deterministic slowly varying potentials
$h_i=h\cos(2\pi\beta i^n+\phi)$ with $0<n<1$~\cite{Sarma1990,Fishman} violate both assumptions in a
controlled way: the potential itself contains only a single global phase $\phi$ --- randomness
enters separately, through the couplings $V_{ij}$ --- and the local level mismatch shrinks
algebraically along the chain, so resonances are supplied by geometry, not by chance.  Two recent studies frame the present work: slowly varying potentials with \emph{short-range}
interactions support locally thermal regions growing with system size~\cite{LiTuDasSarma2025},
and in quasiperiodic chains long-distance correlations carried by \emph{rare} resonant
eigenstates have been identified~\cite{Padhan2026}.  The mechanism below differs from both: the
long-range scale is built from explicit power-law interactions acting on the potential's
deterministic resonance supply, and pair counts near $x_{\rm pair}$ are not rare-event
dominated.  Numerically, this chain with \emph{random} long-range couplings shows a broad nonergodic
regime, a separation between the level-statistics and imbalance scales, and MBL-like behavior
even at $\alpha=0.5$~\cite{yp_nee} --- in tension with the generic counting.

Starting from the microscopic Hamiltonian, we resolve this structure.  The potential alone
fixes the \emph{supply} of local resonances.  The interaction range then reconstructs the
resonant-object ensemble at $\alpha=2n$: below it the objects are asymptotically isolated
two-level bonds; above it, wherever Hartree fragmentation is operative, the diverging
turning-point runs break into multi-site clusters whose intrinsic matching law we find to be
linear, while isolated wing bonds survive at finite density and restore a reduced two-level
logarithm in the full ensemble.  Counting matched pairs coupled at range $R^{-\alpha}$ then
yields the pair-resonance scale $x_{\rm pair}$ at which matched pairs reach order-one
mutual-resonance density --- what we call the pair network, a statement about resonance
density, not about connectivity or delocalization.  The reconstruction changes the local
spectral structure and the matching coefficients but \emph{not} the asymptotic scaling of
$x_{\rm pair}$, which retains the same algebraic exponent and the same logarithm throughout
$0<\alpha<2$; local resonance structure and long-distance resonance proliferation are thus
separately determined.  Independent fluctuations of the coupling coefficients are essential to the long-range
inter-object coupling: spatially uniform interactions undergo a discrete multipole cancellation and couple two
powers faster in $1/R$.  Whether reaching $x_{\rm pair}$ seeds a full delocalization cascade remains an open
dynamical question; what we derive is the scale structure beneath it, which by itself
reorganizes the interpretation of the numerical phenomenology~\cite{yp_nee}.

\begin{figure}[t]\centering
\includegraphics{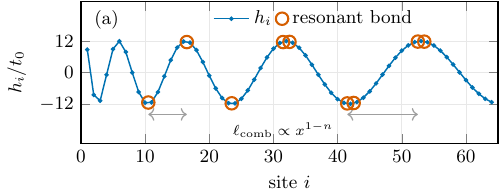}\\[4pt]
\includegraphics{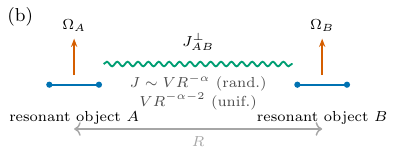}\\[6pt]
\includegraphics{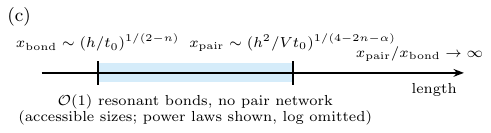}
\caption{Mechanism and scales; panel (c) is schematic.  (a) Resonant bonds --- bonds with $|h_{i+1}-h_i|<t_0$ --- occur only
at the turning points of the slowly varying potential; their spacing grows as
$\ell_{\rm comb}(x)\propto x^{1-n}$.  Shown: $L=64$, $h/t_0=12$, $n=1/2$, with nine resonant
bonds; the phase-averaged supply, Eq.~\eqref{eq:Nres}, predicts $9.33$ against a measured mean of
$9.48$ at this size.  (b) Diagonalizing a local resonant object (bond or island) gives a
pseudospin of splitting $\Omega$; the density interaction couples two such objects a distance $R$
apart through the mixed second difference, Eq.~\eqref{eq:Jzz} --- the bond realization of the
general island form-factor coupling --- surviving as $VR^{-\alpha}$ for random couplings but
cancelling to $VR^{-\alpha-2}$ for uniform ones.  (c) Supply scale and
pair-resonance scale $x_{\rm pair}$ (power-law parts; the relation carries an additional
$\ln x_{\rm pair}$ throughout) separate without bound for $0<\alpha<2$, Eq.~\eqref{eq:separation};
the sizes accessible to the numerical calculations of Ref.~\cite{yp_nee} lie between them.}
\label{fig:mech}
\end{figure}

\emph{Model and resonance supply.}---We consider spinless fermions on an open chain at half
filling,
\begin{equation}
 H=-t_0\sum_{i}\big(c_i^\dagger c_{i+1}+{\rm h.c.}\big)
 +\sum_{i}h_i n_i
 +\sum_{j>i}\frac{V_{ij}}{|i-j|^\alpha}\,n_i n_j,
\label{eq:H}
\end{equation}
with $h_i=h\cos(2\pi\beta i^n+\phi)$, $\beta=(\sqrt5-1)/2$, and independent random couplings
$V_{ij}$ uniform on $[-V,V]$, as in Ref.~\cite{yp_nee}; $n=1/2$ is the principal case and $0<n<1$
is kept general.  Because $[H_{\rm int},n_i]=0$, charge moves only through the nearest-neighbor
hopping; the long-range term transports no charge directly, but after local resonances are
diagonalized it exchanges excitation energy between distant pseudospins~\cite{Serbyn2013,HuseNO2014,Ros2015,Imbrie2016}.

A bond is resonant when the local mismatch $|h_{i+1}-h_i|<t_0$.  The mismatch envelope decays as
$2\pi\beta nh\,i^{n-1}$, so resonant bonds appear only near the turning points of the potential
[Fig.~\ref{fig:mech}(a)], with a resonance probability that grows algebraically along the
chain; phase averaging gives the exact one-bond resonance probability and, in the slowly
varying limit, the asymptotic supply
\begin{equation}
 N_0(L,h)\simeq\frac{1}{\pi^2\beta n(2-n)}\,\frac{t_0}{h}\,L^{2-n},
\label{eq:Nres}
\end{equation}
i.e.\ $N_0\simeq0.219\,(t_0/h)L^{3/2}$ at $n=1/2$; a chain therefore first hosts a resonant
bond at the supply scale $x_{\rm bond}\sim(h/t_0)^{1/(2-n)}$.  Equation~\eqref{eq:Nres} contains no reference
to the interaction, and it is robust: the interaction-induced Hartree shifts $\eta_i$ (below) are
statistically stationary and leave the leading law unchanged~\cite{companion}.  The supply of
local resonances is fixed by the potential alone.

\emph{The local resonant object.}---What \emph{is} a resonance here?  Near a turning point at
position $x$ the mismatch stays below $t_0$ over a run of
$w_0(x)=[2t_0/(2\pi\beta n)^2]\,x^{2-2n}/h$ consecutive bonds, which grows with $x$ at fixed $h$.
The relevant question is how $w_0$ behaves \emph{along the trajectory} $h\propto
x_{\rm pair}^{(4-2n-\alpha)/2}$ derived below, where
\begin{equation}
 w_0(x_{\rm pair})\sim\sqrt{\frac{t_0}{V}}\;
 \frac{x_{\rm pair}^{(\alpha-2n)/2}}{(\ln x_{\rm pair})^{1/2}},
 \qquad 0<\alpha<2,
\label{eq:w0traj}
\end{equation}
Because the matching logarithm derived below is present throughout $0<\alpha<2$
[Eq.~\eqref{eq:Pmix}], Eq.~\eqref{eq:w0traj} holds without case distinction, and the exponent
changes sign at $\alpha=2n$.  For $\alpha<2n$, $w_0\to0$: resonances are asymptotically isolated
two-site bonds --- two-level systems (TLS) with splitting $\Omega=\sqrt{\Delta^2+4t_0^2}$ and a
hard band edge at $2t_0$ shared by every bond.  At $\alpha=2n$ the logarithm decides,
$w_0\sim(\ln x_{\rm pair})^{-1/2}\to0$, so the equality is asymptotically bond-like, though only
logarithmically so; for $n=1/2$ this places $\alpha=1$ on the bond side.  For $\alpha>2n$ the
bare runs grow without bound, a purely two-level description of the turning-point region fails,
and the resonant-object ensemble is decided by the Hartree field.

\begin{figure}[t]\centering
\includegraphics{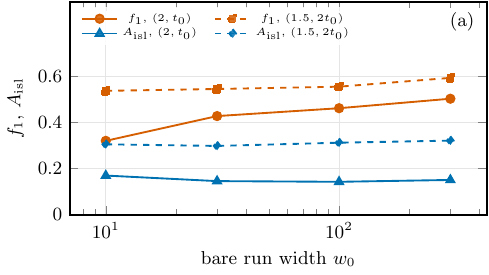}\\[3pt]
\includegraphics{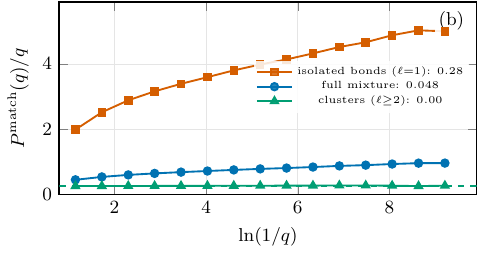}
\caption{Hartree fragmentation changes the matching ensemble.  (a) After recursive boundary
healing, the single-bond fraction $f_1$ of active objects shows no downward trend and the
active-object density $A_{\rm isl}=\rho_{\rm act}/p$ is $w_0$-independent [Eq.~\eqref{eq:f1}].
(b) Matching law by sector at $\alpha=2$, $V=t_0$, from island diagonalization: $\ell\ge2$
clusters are flat, surviving $\ell=1$ objects retain a finite logarithmic slope, and the full
ensemble slope is $b_{\rm all}\simeq f_1^2b_1$ ($0.41^2\times0.28=0.047$ versus the measured
$0.048$).  Statistical errors: $\pm0.01$ on $f_1$ and $A_{\rm isl}$ (batch means over
$48$ chains); fitted slopes $\pm0.01$.  Protocols and the remaining $(\alpha,V)$ sets in the
Supplemental Material~\cite{SMletter}.}
\label{fig:frag}
\end{figure}

\emph{Hartree fragmentation.}---The physical detuning of bond $i$ is
$\Delta_i=(h_{i+1}-h_i)+\eta_i$, where
$\eta_i=\sum_{j\neq i,i+1}(U_{i+1,j}-U_{i,j})\,n_j$ is the random Hartree mismatch --- the
exact diagonal interaction contribution to the detuning of a Fock-state hop, not a mean-field
approximation --- a stationary, summably correlated sequence.  Near the centre of a bare run
the deterministic mismatch vanishes, so the run is cut wherever $|\eta_i|>t_0$, with
probability $p_{\rm br}(V,\alpha)=\Pr(|\eta_i|>t_0)$ per bond; runs therefore fragment on the
characteristic scale
\begin{equation}
 \xi_{\rm frag}=p_{\rm br}^{-1},
\label{eq:xifrag}
\end{equation}
independent of position and scale (the measured island sizes track $\xi_{\rm frag}$ to
within $10$--$20\%$~\cite{SMletter}; $\eta_i$ is correlated, so this is a characteristic scale,
not an independent-break island length).  But $p_{\rm br}$ can vanish identically: for $\alpha>1$
the Hartree field has strictly bounded support, ${\rm ess\,sup}\,|\eta_i|=S(\alpha)V$ with
$S(\alpha)=2[2\zeta(\alpha)-1]$, so $p_{\rm br}=0$ below the support threshold
$V_*(\alpha)=t_0/S(\alpha)$ [$0.118\,t_0$ at $\alpha=3/2$, $0.218\,t_0$ at $\alpha=2$].
Fragmentation is operative only when $p_{\rm br}>0$ \emph{and} $w_0\gg\xi_{\rm frag}$; just above
$V_*$ it is large-deviation suppressed ($\xi_{\rm frag}\gtrsim10^6$ at $\alpha=1.25$,
$V=0.25\,t_0$).  Along the $\alpha>2n$ trajectory $w_0\to\infty$, so any $p_{\rm br}>0$ eventually
fragments every run.  This island classification is established for $n\ge1/4$ --- including the
principal case $n=1/2$ --- where the fragmented branch has $\alpha>1/2$ and the Hartree width is
scale independent; in the corner $n<1/4$, $2n<\alpha\le1/2$, the Hartree width grows with
system size, so $p_{\rm br}$ is not a fixed number and the island classification becomes
scale dependent; that corner is outside the present theory~\cite{companion}.  Fragmentation does not, however, convert the ensemble into a pure island gas.  In the scaled wing
coordinate $y=2m/w_0$, where $m$ is the offset from the turning point, the bare mismatch is
$\delta^{(0)}/t_0\simeq-y$ and the resonance condition reads $|\eta/t_0-y|<1$.  Since $\eta$ is
stationary with a $w_0$-independent distribution, both the resonance probability
$p_{\rm res}(y)$ and the singleton probability
$p_{\rm sing}(y)=\Pr(\chi_i=1,\chi_{i\pm1}=0\mid y)$ are $\order(1)$ functions of $y$ alone,
positive on a finite interval --- for the latter because adjacent Hartree shifts are
anticorrelated rather than locked~\cite{SMletter}.  An interval $\dd y$ holds $(w_0/2)\dd y$
bonds, so the wings contribute $\propto w_0$ isolated objects, exactly as the fragmented interior
contributes $\propto w_0$ islands.  Their ratio therefore does not vanish: the fraction of active
objects that are single bonds tends to a constant,
\begin{equation}
 f_1(w_0)\ \longrightarrow\ f_1^{\infty}=\order(1),
\label{eq:f1}
\end{equation}
measured as $f_1=0.32$--$0.50$ at $(\alpha,V)=(2,t_0)$ and $0.54$--$0.59$ at $(1.5,2t_0)$ across
$w_0=10$--$300$, with no downward trend [Fig.~\ref{fig:frag}(a)] --- and unchanged when the
partition is iterated to a fixed point against the islands' own internal
rearrangements~\cite{SMletter}.  The active-object density $A_{\rm isl}=\rho_{\rm act}/p$ is
likewise $w_0$-independent.  Crossing $\alpha=2n$ therefore \emph{reconstructs} the
resonant-object ensemble --- from asymptotically bond-like below, to a mixture of
Hartree-fragmented multi-site clusters and surviving isolated wing bonds above --- rather than
replacing one elementary object by another.  What is $w_0$-independent above $\alpha=2n$ is the
density of \emph{typical} $\order(1)$ clusters; a rare, bond-weighted tail does broaden under
recursive healing; a transition-weight sum rule keeps its total contribution per unit length
bounded, and the measured low-frequency matching density shows no growth over the cluster sizes
reached numerically~\cite{companion}.

\emph{Matching in the fragmented ensemble.}---A pair of resonant objects at distance $R$ forms
a resonant \emph{pair} when their splittings match to within the coupling.  For bonds the
resonance condition carries the mixing angles,
$|\Omega_A-\Omega_B|<|J^{zz}_{AB}|\,s_As_B$ with $s_i=2t_0/\Omega_i$, and the dimensionless
window is $q\equiv|J^{zz}|/t_0$.  For isolated bonds the splitting distribution has an integrable
singularity at the shared band edge $\Omega_{\rm min}=2t_0$; matching two independent draws then
gives the logarithmically enhanced law $P^{\rm match}=A\,q\ln(1/q)+\order(q)$, with $A=2$ in
this convention --- verified against direct ensemble counting with no free
parameter~\cite{SMletter}.  Fragmented islands share no
deterministic band edge: each island's lowest splitting is set by an island-specific avoided
crossing.  Exact diagonalization of islands harvested from the full-detuning statistics (up to
$\ell=20$ bonds~\cite{SMletter}) shows that the pair-density diagnostic
$D(w)=\Pr(|\Omega_A-\Omega_B|<w)/2w$ saturates over the accessible window, i.e.\ the intrinsic
island law is linear over the resolved window [Fig.~\ref{fig:frag}(b), green]:
\begin{equation}
 P^{\rm match}\simeq C_{\rm isl}\,\frac{|J^{\perp}|}{t_0},\qquad 10^{-4}\lesssim |J^\perp|/t_0\lesssim10^{-1},
\label{eq:Pisl}
\end{equation}
with $C_{\rm isl}$ an $\order(1)$, island-dependent constant (the window convention absorbed
into it).  Two limitations are stated here rather than left to the Supplemental Material: the
recursively healed ensemble has a broad cluster tail, most of whose resonant-bond weight at
$(\alpha,V)=(2,t_0)$ lies in clusters larger than those diagonalized, and the absence of a
low-frequency singularity in that tail is an assumption supported by, not derived from, the
data~\cite{SMletter}.  The two-bond ($\ell=2$) population is likewise slope-free,
and the saturation persists when \emph{all} island transitions with nonvanishing local transition
density are included at infinite-temperature weighting~\cite{SMletter}.  Fragmentation therefore removes the two-level matching singularity
from the multi-site objects themselves.

Because $f_1$ remains finite, the physical ensemble is a mixture.  With $w_{11}$, $w_{1I}$,
$w_{II}$ the pair-sector weights on a shell,
\begin{equation}
\begin{aligned}
 P^{\rm match}&=w_{11}P_{11}+w_{1I}P_{1I}+w_{II}P_{II}\\
 &\simeq b_{\rm all}\,q\ln(1/q)+C_{\rm all}\,q,
\end{aligned}
\label{eq:Pmix}
\end{equation}
with $b_{\rm all}=w_{11}b_1>0$: only bond--bond pairs share the deterministic edge, so only
they produce a logarithm, but they do so at finite weight.  The comb correlates objects, so
$w_{11}$ need not equal $f_1^2$; measured on the physical mesoscopic shell it does to within
$3$--$6\%$~\cite{SMletter}, and $b_{\rm all}\simeq f_1^2b_1$ holds both for the snapshot
partition and at the healed fixed point~\cite{SMletter}.  Measured,
$b_{\rm all}=0.048$--$0.24$ against the analytic isolated-bond value $A=2$ --- a suppression by a
factor of $8$--$40$ --- while $C_{\rm all}\simeq0.5$--$0.8$.  Since $q\ln(1/q)\gg q$ as
$q\to0$, the logarithmic channel nevertheless dominates beyond $R\gtrsim10$--$300$ sites, far
inside $x_{\rm pair}$.  The change at $\alpha=2n$ thus alters the local spectral structure and
the coefficients, \emph{not} the asymptotic form of the matching law.

\begin{figure}[t]\centering
\includegraphics{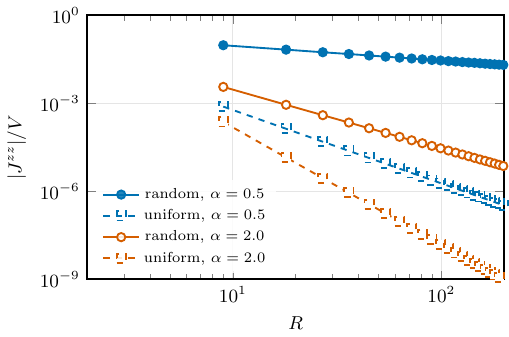}
\caption{Independent coefficient fluctuations are essential.  Effective two-bond coupling at identical bare strength $V$:
random coefficients (circles) leave the mixed second difference of Eq.~\eqref{eq:Jzz}
uncancelled, shown as the rms $VR^{-\alpha}/(2\sqrt3)$; uniform coefficients (squares) cancel to
the deterministic value $-\tfrac{V}{4}\alpha(\alpha+1)R^{-\alpha-2}$, shown as its magnitude.
For $\alpha\le1$ the uniform fixed-$V$ control is not Kac normalized; the comparison is between
local resonant matrix elements, for which the cancellation is unchanged.  The gap reaches four orders of magnitude by
$R=100$ at $\alpha=0.5$.}
\label{fig:randunif}
\end{figure}

\emph{Long-range coupling: random versus uniform.}---For two bonds $A=(a,a{+}1)$, $B=(b,b{+}1)$ at
distance $R$, projecting the density interaction onto the pseudospins leaves exactly the mixed
second difference
\begin{equation}
 J_{AB}^{zz}=\tfrac14\big(U_{ab}-U_{a,b+1}-U_{a+1,b}+U_{a+1,b+1}\big),
\label{eq:Jzz}
\end{equation}
$U_{ij}=V_{ij}/|i-j|^\alpha$.  For independent random couplings nothing cancels:
$J^{zz}_{\rm rms}=VR^{-\alpha}/(2\sqrt3)$.  For uniform couplings $U_{ij}=V/|i-j|^\alpha$ the
same expression is a smooth second difference,
$J^{zz}_{AB,\rm unif}\simeq-\tfrac{V}{4}\alpha(\alpha+1)R^{-\alpha-2}$ --- two powers faster, a
discrete multipole cancellation of the type familiar from long-range localization
theory~\cite{Yao2014}, here derived as the microscopic control channel
[Fig.~\ref{fig:randunif}].  For islands the same structure appears with transition-density form
factors, $J^\perp_{AB}=\sum_{a\in A,b\in B}U_{ab}m^A_am^B_b$: finite island size renormalizes the
\emph{amplitude} by $\order(1)$ factors while the random-coupling \emph{power} $R^{-\alpha}$ is
untouched.  The uniform cancellation also survives fragmentation exactly: in a fixed-number
sector both island monopoles vanish, $\sum_am^A_a=\langle g_A|N_A|e_A\rangle=0$, so the uniform
island coupling starts at the dipole--dipole order,
$J^{\perp}_{\rm unif}\propto VR^{-\alpha-2}$~\cite{companion}.  The operative property is the
independence of the coefficients entering the mixed second difference; spatially correlated
couplings would partially restore the cancellation.  This is the microscopic reason random and uniform long-range interactions behave
qualitatively differently in this model~\cite{yp_nee} and in closely related models~\cite{Nag2019}.

\emph{The pair-resonance scale.}---Because every object is a deterministic function of the single
phase $\phi$, occurrence and matching are controlled by the same comb and the pair density cannot
be factorized into marginals.  The construction instead coarse grains over a \emph{mesoscopic
shell}, $\ell_{\rm comb}(x)\ll\Delta R\ll x$ --- a window that exists asymptotically because
$\ell_{\rm comb}/x\sim x^{-n}$.  Across it the comb offset sweeps $\order(1)$ of phase while the
envelopes and the coupling vary by $\order(\Delta R/x)$, so the shell average of the \emph{full}
joint resonance-and-matching indicator (not of its factors separately) is smooth.  This coarse
graining is analytically motivated and numerically validated rather than proved: the
shell-averaged correlation factor is measured to be unity within errors, and at finite $V$ the
Hartree field independently supplies a continuous joint measure~\cite{companion}.  It returns the
scaling form
$n_R(x)\sim R\,\rho(x)\rho(x-R)\,\langle P^{\rm match}(q_R)\rangle$, $q_R\sim V/(t_0R^\alpha)$,
with $\rho=p(x)\propto(t_0/h)x^{1-n}$ the resonant-bond density below $\alpha=2n$ and
$\rho_{\rm act}=A_{\rm isl}p$, $A_{\rm isl}=\order(1)$, the active-object density above
it~\cite{companion}.  Because both sectors of Eq.~\eqref{eq:Pmix} are present with finite weight
throughout, the chain-integrated pair count carries both channels,
\begin{equation}
\begin{aligned}
 \Lambda(x)&\sim\frac{Vt_0}{h^2}\,x^{\,4-2n-\alpha}\big[B\ln x+C\big],\\
 &\qquad B\propto b_{\rm all}=w_{11}b_1>0,
\end{aligned}
\label{eq:Lambda}
\end{equation}
and $\Lambda\sim1$ --- matched pairs at order-one mutual-resonance density --- gives a
\emph{single} law for $x_{\rm pair}$ on both sides,
\begin{equation}
 x_{\rm pair}^{\,4-2n-\alpha}\,\ln x_{\rm pair}\propto \frac{h^2}{Vt_0},
 \qquad 0<\alpha<2 .
\label{eq:xpair}
\end{equation}
The elementary objects change at $\alpha=2n$; the scaling of $x_{\rm pair}$ does not.  On the
fragmented side this statement is conditional on the two limitations stated after
Eq.~\eqref{eq:Pisl} --- a bounded low-frequency matching density along the rare cluster tail,
and the shell coarse graining of the deterministic comb, which is numerically validated rather
than proved (below).  A direct fixed-shell test supports this scaling within the
resonance-starved regime: rescaling the measured pair count by $h^2x^{-(4-2n-\alpha)}/(Vt_0)$
gives the prescribed algebraic dependence at $\alpha=0.5$, $1$, and $1.5$, with no systematic
deterioration on the fragmented side; the remaining finite-range deviations are dominated by the
active-object density, and the accessible $\ln x$ interval does not distinguish the predicted
logarithm from a more general slowly varying correction [Fig.~\ref{fig:lamtest}; full test in
the Supplemental Material~\cite{SMletter}].  If system size is the limiting scale,
$x_{\rm pair}\sim L$ converts Eq.~\eqref{eq:xpair} into a size-dependent pair-resonance scale
\begin{equation}
 h_{\rm pair}(L)\propto\sqrt{Vt_0}\;L^{(4-2n-\alpha)/2}\,\sqrt{\ln L},
\label{eq:hcapp}
\end{equation}
with exponent $(3-\alpha)/2$ at $n=1/2$.  Comparing with the supply scale
$x_{\rm bond}\sim(h/t_0)^{1/(2-n)}$,
\begin{equation}
 \frac{x_{\rm pair}}{x_{\rm bond}}\sim
 \Big(\frac{h}{t_0}\Big)^{\!\alpha/[(2-n)(4-2n-\alpha)]}
 \Big(\frac{V}{t_0}\Big)^{\!-1/(4-2n-\alpha)}\longrightarrow\infty
\label{eq:separation}
\end{equation}
up to a slowly varying $[\ln(h^2/Vt_0)]^{-1/(4-2n-\alpha)}$ factor from the logarithm of
Eq.~\eqref{eq:xpair}, dropped here because only the comparison of powers matters.  The exponent
of $h$ is positive for every fixed $0<\alpha<2$, so the two scales separate without bound in the
strong-potential limit at fixed $V/t_0$.  At $\alpha=2$ the two marginalities meet --- the shell measure
$\dd\kappa/\kappa$ contributes one logarithm and the surviving matching channel of
Eq.~\eqref{eq:Pmix} another --- so that
$x_{\rm pair}^{2(1-n)}(\ln x_{\rm pair})^2\propto h^2/(Vt_0)$, i.e.\
$x_{\rm pair}(\ln x_{\rm pair})^2\propto h^2/(Vt_0)$ at $n=1/2$; the separation persists there
too~\cite{companion}.

Are pair counts near order-one mean carried by rare configurations?  We count matched pairs on
a fixed mesoscopic shell, at the island level (frozen islands excluded), for representative
parameters near $x_{\rm pair}$ at $\alpha=0.5$, $1$, and $1.5$ --- i.e.\ on both sides of the reconstruction and at
the marginal point.  At matched mean $\langle N_{\rm pair}\rangle\simeq1.4$ the three distributions
are closely comparable: median $1$, $P(N_{\rm pair}\ge1)=0.63$--$0.69$ against the Poisson value
$0.75$--$0.76$, with variance $1.6$--$1.8$ times Poisson (Table~\ref{tab:typ}) --- modestly
overdispersed, but rare-event dominated in none of the three cases; the result is essentially
unchanged when the Gaussian amplitude of the matching window is replaced by the physical island
form-factor coupling~\cite{SMletter}.  This tests the sample statistics at matched mean, not the
location or scaling of $x_{\rm pair}$ itself.

\begin{figure}[t]\centering
\includegraphics{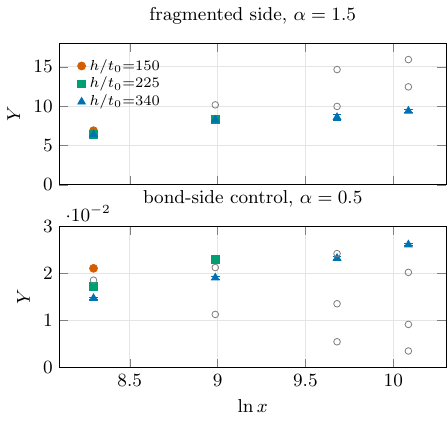}
\caption{Direct fixed-shell test of the pair-count scaling, Eq.~\eqref{eq:Lambda}:
$Y\equiv\Lambda\,h^2x^{-(4-2n-\alpha)}/(Vt_0)$ on a shell at $R=0.45\,x$, with the algebraic
exponent \emph{fixed by Eq.~\eqref{eq:Lambda}, not fitted}.  Colored symbols: points inside the
resonance-starved window $w_0/\ell_{\rm comb}\le0.15$; gray open symbols: outside it, where the
object-density asymptotics fail~\cite{SMletter}.  Within the window the $h^{-2}$ rescaling
collapses the series and the fragmented side shows no progressive failure relative to the
bond-side control; the accessible range tests the predicted algebraic scaling and its
persistence across the reconstruction, and does not independently resolve the logarithmic
correction.}
\label{fig:lamtest}
\end{figure}

\emph{Discussion.}---The derived scale structure reorganizes the numerical phenomenology of this
model~\cite{yp_nee}.  Accessible system sizes sit \emph{between} the two scales
[Fig.~\ref{fig:mech}(c)]: chains large enough to contain a handful of resonant bonds but far too
small to contain a pair network.  The supply law Eq.~\eqref{eq:Nres} places an order-one
forward-branching scale on the resonant Fock-space graph at $h_{\rm FB}\propto L^{2-n}$ --- and
exact construction of that graph at $L\le18$ shows no giant component there (End Matter) --- while
the pair-resonance scale Eq.~\eqref{eq:hcapp} grows with the parametrically larger exponent
$(4-2n-\alpha)/2$.  This two-scale structure provides a natural
interpretation of the separation the numerics reported between the level-statistics crossover
$h_c^r$ and the imbalance scale $h_c^I$~\cite{yp_nee}, while leaving the thermodynamic status
of the intermediate regime open~\cite{Suntajs2020,Tu2023}; the anomalously slow dynamics of
the intermediate window is consistent with sparse $\order(1)$ resonant objects coupled at
$VR^{-\alpha}$ below $x_{\rm pair}$.  The random-versus-uniform contrast [Fig.~\ref{fig:randunif}]
explains why uniform long-range interactions behave as effectively short-ranged here.  The
calculation does not determine whether the order-one pair density at $x_{\rm pair}$ seeds an
all-generation cascade to delocalization, or whether higher-generation resonances remain
starved.  That question and the spectral signatures of the two scales are the natural next
targets, most directly testable through the random-versus-uniform contrast at fixed $V$ and
through spatially resolved pseudospin-versus-charge memory~\cite{Schreiber2015,Smith2016}.

\begin{acknowledgments}
\emph{Acknowledgments.}---This work is supported by the National Research Foundation of Korea
(NRF), funded by the Ministry of Science and ICT (MSIT), Korea, under
Grant Nos.~NRF-2022R1A2C1011646, RS-2024-00416036, RS-2025-03392969,
and RS-2026-25490114; by the Creation of the Quantum Information
Science R\&D Ecosystem program through the NRF funded by the Korean
government (MSIT) (Grant No.~RS-2023-NR068116); by the Quantum
Simulator Development Project for Materials Innovation through the NRF
funded by MSIT, Korea (Grant No.~RS-2023-NR119931); by the Brain Pool
Program funded by MSIT through the NRF (Grant No.~RS-2025-25446099);
and by the Institute of Information \& Communications Technology
Planning \& Evaluation (IITP) through the Information Technology
Research Center (ITRC) program, funded by MSIT, under the project
Quantum--AI Convergence for Innovation and Talent Development
(Grant No.~RS-2026-25519864).
\end{acknowledgments}

\emph{Data availability.}---The numerical data supporting this article, together with the
analysis and simulation code used to generate the reported protocols and figures, are available
from the contact author upon reasonable request.

\bibliography{TwoScales}

\section*{End Matter}

\emph{Sample statistics near $x_{\rm pair}$.}---Islands are harvested from chains of $8000$
bonds at $h=150\,t_0$, $V=t_0$, with the full detuning, centered at $x_0=4000$; frozen islands
are excluded, and each active island contributes one node with its exact doublet
$(\Omega,\lVert m\rVert)$.  Matched pairs between the center window and a shell at
$\bar R=1800$ (width $4\ell_{\rm comb}$) are counted, with matching window
$2\lambda(V/R^\alpha)|g|\,\lVert m^A\rVert\lVert m^B\rVert$, $g$ Gaussian; $\lambda$ is an
auxiliary rescaling comparing the distributions at matched mean across $\alpha$ (it absorbs
the fixed-shell $R^{-\alpha}$ factor), not a physical coupling prefactor.  Table~\ref{tab:typ} reports the $\lambda$ giving $\langle N\rangle\simeq1.4$ in
each regime ($600$ realizations); full scans, the physical-coupling repeat, and a
one--two-bond repeat ($w_0\simeq1.9$) give closely comparable distributions~\cite{SMletter}.

\begin{table}[h]
\caption{Shell pair-count distribution at matched mean, at the island level, across the
object reconstruction ($n=1/2$: $\alpha=0.5$ bond side, $\alpha=1$ marginal, $\alpha=1.5$
fragmented regime; $V=t_0$, $w_0\simeq14$).  Poisson reference values in brackets.}
\label{tab:typ}
\begin{ruledtabular}
\begin{tabular}{lcccc}
$\alpha$ & $\langle N\rangle$ & median & $P(N\ge1)$ & ${\rm Var}/\langle N\rangle^2$\\
\colrule
$0.5$ & $1.44$ & $1$ & $0.69$ [$0.76$] & $1.13$ [$0.70$]\\
$1.0$ & $1.41$ & $1$ & $0.64$ [$0.75$] & $1.24$ [$0.71$]\\
$1.5$ & $1.38$ & $1$ & $0.63$ [$0.75$] & $1.23$ [$0.73$]\\
\end{tabular}
\end{ruledtabular}
\end{table}

\emph{Fragmentation constants.}---$p_{\rm br}(V,\alpha)=\Pr(|\eta|>t_0)$ is sampled from the
exact Hartree ensemble at half filling~\cite{SMletter}.  For $\alpha>1$ the support bound
${\rm ess\,sup}|\eta|=S(\alpha)V$, $S(\alpha)=2[2\zeta(\alpha)-1]$, gives the threshold
$V_*=t_0/S(\alpha)$ below which $p_{\rm br}=0$ identically; for $\alpha\le1$ the column sums diverge
with $L$, so $p_{\rm br}>0$ for any $V>0$ at large enough $L$ (Table~\ref{tab:frag}).

\begin{table}[h]
\caption{Support threshold $V_*$ and breaking probability $p_{\rm br}$ (half filling);
entries $<5\times10^{-7}$: large-deviation suppression just above $V_*$.}
\label{tab:frag}
\begin{ruledtabular}
\begin{tabular}{lccccc}
$\alpha$ & $V_*/t_0$ & \multicolumn{4}{c}{$p_{\rm br}$ at $V/t_0=$}\\
 & & $0.25$ & $0.5$ & $1$ & $2$\\
\colrule
$1.25$ & $0.061$ & $<5\times10^{-7}$ & $0.010$ & $0.180$ & $0.486$\\
$1.5$  & $0.118$ & $<5\times10^{-7}$ & $0.005$ & $0.146$ & $0.442$\\
$2$    & $0.218$ & $<5\times10^{-7}$ & $0.002$ & $0.113$ & $0.402$\\
\end{tabular}
\end{ruledtabular}
\end{table}

\emph{Forward branching.}---The supply law Eq.~\eqref{eq:Nres} places mean forward branching
$B=1$ of the resonant Fock graph at $h_{\rm FB}\propto L^{2-n}$; exact construction at $L\le18$
finds the crossing at $N_{\rm res}=2.4$--$2.7$ and no giant component --- a finite-size
control, not percolation~\cite{SMletter,companion}.

\clearpage
\onecolumngrid
\begin{center}\textbf{\large Supplemental Material for ``Hartree Fragmentation and Long-Range Pair Networks in Aperiodic Chains''}\end{center}
\vspace{2mm}\twocolumngrid
\setcounter{equation}{0}\setcounter{figure}{0}\setcounter{table}{0}\setcounter{section}{0}
\providecommand{\e}{\mathrm{e}}\providecommand{\ii}{\mathrm{i}}\providecommand{\dd}{\mathrm{d}}
\renewcommand{\theequation}{S\arabic{equation}}\renewcommand{\thefigure}{S\arabic{figure}}\renewcommand{\thetable}{S\Roman{table}}

\noindent This Supplemental Material collects the numerical protocols behind every number quoted
in the Letter: the Hartree-fragmentation statistics --- breaking probability, support threshold,
and island sizes (Sec.~S1); the island-diagonalization protocol and the matching-law
diagnostic of Fig.~2(b) (Sec.~S2); the island-level shell pair-count typicality
scans at $\alpha=0.5$, $1$, and $1.5$ behind Table~I (Sec.~S3); the exact resonant
Fock-graph protocol behind the End Matter (Sec.~S4); the analytic isolated-bond
matching benchmark (Sec.~S5); and the direct fixed-shell test of the pair-count scaling
(Sec.~S6).  Extended derivations --- Hartree covariance identities,
the shell-equidistribution analysis, conditional resonance distributions, and the marginal
$\alpha=2$ construction --- are given in Ref.~\cite{companion}.  Equation, figure,
and table numbers without an ``S'' refer to the Letter.

\section*{S1.\ Hartree fragmentation: breaking probability, support threshold, and island statistics}

\subsection*{Breaking probability and support threshold}

The breaking probability $p_{\rm br}(V,\alpha)=\Pr(|\eta_i|>t_0)$ is evaluated by direct sampling
of the Hartree detuning $\eta_i=\sum_{j\neq i,i+1}(U_{i+1,j}-U_{i,j})n_j$ over $2\times10^6$
independent draws of the couplings and half-filled occupations, with the spatial sums cut at
range $2000$ (the omitted tail is negligible for the $\alpha$ values shown).  For $\alpha>1$ the
support of $\eta_i$ is strictly bounded, $\operatorname{ess\,sup}|\eta_i|=S(\alpha)V$ with
$S(\alpha)=2[2\zeta(\alpha)-1]$, giving the support threshold $V_*(\alpha)=t_0/S(\alpha)$ of the
Letter (Table~\ref{smtab:pbr}, an expanded version of Table~II of the Letter).  An independent
measurement of the same quantity inside the turning-point windows of the island ensemble below
agrees to $\sim10\%$ (e.g.\ $0.0058$, $0.147$, and $0.437$ at $V/t_0=0.5$, $1$, and $2$,
respectively, at $\alpha=1.5$), so
$\xi_{\rm frag}=p_{\rm br}^{-1}$ therefore provides a quantitative characteristic
fragmentation scale, tracking the measured snapshot island scale at the $10$--$20\%$ level.

\begin{table}[h]
\caption{Breaking probability $p_{\rm br}$ and support threshold.  Entries $<5\times10^{-7}$ are
unresolved at the sampling depth; for $\alpha=1.25$, $V=0.25\,t_0$ the ratio $V/V_*\simeq4.1$, so
$p_{\rm br}>0$ strictly, yet $\xi_{\rm frag}=p_{\rm br}^{-1}\gtrsim10^6$ --- fragmentation is
mathematically allowed but operationally absent, which is why the controlled-domain condition is
$w_0\gg\xi_{\rm frag}$ and not merely $V>V_*$.}
\label{smtab:pbr}
\begin{ruledtabular}
\begin{tabular}{ccccccc}
$\alpha$ & $S(\alpha)$ & $V_*/t_0$ & \multicolumn{4}{c}{$p_{\rm br}$ at $V/t_0=$}\\
 & & & $0.25$ & $0.5$ & $1$ & $2$\\
\colrule
$1.25$ & $16.38$ & $0.061$ & $<5{\times}10^{-7}$ & $0.0101$ & $0.180$ & $0.486$\\
$1.50$ & $8.45$  & $0.118$ & $<5{\times}10^{-7}$ & $0.0051$ & $0.146$ & $0.442$\\
$2.00$ & $4.58$  & $0.218$ & $<5{\times}10^{-7}$ & $0.0021$ & $0.113$ & $0.402$\\
\end{tabular}
\end{ruledtabular}
\end{table}

\subsection*{Island-size statistics}

Islands are maximal runs of consecutive bonds with $|\Delta_i|<t_0$, $\Delta_i$ the full detuning
(deterministic mismatch plus Hartree field).  The ensemble uses chains of $4\times10^4$ bonds
centered at $x_0=10^6$, the exact potential at $n=1/2$, $\beta=(\sqrt5-1)/2$, random couplings
with range cutoff $300$, random half-filled occupations, and a fresh global phase per realization
($100$ realizations per parameter set).  The field $h$ is chosen to set the bare run width
$w_0=2t_0/(h\psi'^2)$; increasing $w_0$ at fixed $V$ is equivalent to moving outward along the
pair-network trajectory.  Table~\ref{smtab:isl} gives the bond-weighted mean island size
$\langle\ell\rangle_B=\langle\ell^2\rangle/\langle\ell\rangle$; the underlying $\alpha=2$ ensemble is the one from which
Fig.~2(a) of the Letter derives the single-bond fraction and active-object density.  In every parameter set the resonant-bond density agrees with the
bare density to within $0.3\%$, and the count-weighted mean $\langle\ell\rangle_{\rm isl}$ is
much smaller than $\langle\ell\rangle_B$ (e.g.\ $2.8$ against $7.6$ at $\alpha=2$, $V=t_0$,
$w_0=100$): the island-size distribution is broad, and the tail carries the bond weight.

\begin{table}[h]
\caption{Bond-weighted mean island size $\langle\ell\rangle_B$ against bare run width $w_0$.
Saturation sets in when $w_0\gg\xi_{\rm frag}$, at a level consistent with $\xi_{\rm frag}$:
$6.8$ and $2.3$ at $\alpha=1.5$, and $8.8$ and $2.5$ at $\alpha=2$, for $V/t_0=1$ and $2$,
respectively.}
\label{smtab:isl}
\begin{ruledtabular}
\begin{tabular}{cccccc}
$\alpha$ & $V/t_0$ & $w_0{=}10$ & $30$ & $100$ & $300$\\
\colrule
$1.5$ & $0.25$ & $9.40$ & $24.3$ & $64.7$ & $155.6$\\
$1.5$ & $0.5$  & $7.54$ & $15.3$ & $28.8$ & $44.8$\\
$1.5$ & $1$    & $4.49$ & $5.83$ & $6.27$ & $6.25$\\
$1.5$ & $2$    & $2.35$ & $2.43$ & $2.45$ & $2.33$\\
$2$   & $0.25$ & $9.59$ & $25.1$ & $68.9$ & $167.9$\\
$2$   & $0.5$  & $7.97$ & $17.0$ & $33.9$ & $59.1$\\
$2$   & $1$    & $5.00$ & $6.66$ & $7.60$ & $7.65$\\
$2$   & $2$    & $2.59$ & $2.69$ & $2.75$ & $2.63$\\
\end{tabular}
\end{ruledtabular}
\end{table}

\subsection*{Boundary stability of the island partition}

The islands are harvested from a reference occupation configuration; internal hopping changes
occupations and thereby the Hartree detunings of the boundary (cut) bonds.  We test whether the
partition survives its own dynamics.  For every cut bond between two adjacent islands (both
$\ell\le6$ bonds; larger neighbors, $2$--$27\%$ of cuts depending on $w_0$, are skipped), all
internal Fock configurations of \emph{both} adjacent islands (fixed particle number each) are
enumerated, the cut bond's spectator-only Hartree detuning is recomputed for each configuration
pair, and the cut is called \emph{healable} if any configuration makes it active (resonant,
$|\Delta_{\rm cut}|<t_0$, and flippable).  This is the most permissive rule --- any single
configuration suffices.  Two realizations $\times$ six phases of $4\times10^4$-bond chains per
parameter set; $w_0\in\{10,30,100,300\}$.

The healing probability per cut is substantial but \emph{$w_0$-independent} --- e.g.\
$P_{\rm heal}=0.65/0.61/0.58/0.57$ at $(\alpha,V)=(2,t_0)$ and $0.36$--$0.41$ at $(1.5,2t_0)$
for $w_0=10/30/100/300$ --- and the fraction of configuration pairs that heal a given cut is
small and flat ($f_{\rm heal}\simeq0.09$--$0.18$).  Merging islands across every healable cut
(the most pessimistic aggregation), the bond-weighted merged-cluster size saturates in $w_0$:
$7.1/9.2/9.8/10.2$ bonds at $(2,t_0)$ against the unhealed $\langle\ell\rangle_B\simeq7.6$, and
$4.0$--$5.1$ at $(2,2t_0)$.  Healing therefore re-draws boundaries locally and renormalizes the
island scale by an $\order(1)$ factor absorbed in the nonuniversal island constants.

\emph{Recursive closure.}  A merged cluster can in principle reopen further cuts, so we iterate
the merge to a fixed point, re-evaluating the remaining cuts against the \emph{merged} clusters.
To leave no enumeration gap, each cluster is represented at a cut by the $w\le8$ sites nearest
it.  This truncation is validated by convergence, not by a decay bound: since the $V_{ij}$ are
independent, $g_j=\order(|j-c|^{-\alpha})$ realization by realization (the smooth-coupling
$|j-c|^{-\alpha-1}$ estimate would contradict the multipole mechanism of the Letter), and the
worst-case remote influence is not small ($0.22\,t_0$ and $2.65\,t_0$).  The omitted tail is
small only in the ensemble sense (rms $0.009\,t_0$ and $0.064\,t_0$); with no analytic tail added,
$w_{\rm cut}=4,8,10$ give the same observables to better than $1.5\%$ (e.g.\ $A_{\rm isl}=
0.1476,0.1455,0.1454$ at $(2,t_0)$, $w_0=30$).  Clusters fitting
the window are enumerated at fixed particle number; a window inside a larger cluster may
exchange one particle with the remainder; a cut reopens if \emph{any} admissible configuration
returns it to the window, with no energetic accessibility imposed, so the rule over-counts
reopening.  (Validation: first-pass probabilities reproduce the exact fixed-number values,
$0.66/0.61/0.59$ versus $0.65/0.61/0.58$ at $(2,t_0)$.)

The recursion converges in $4$--$7$ passes.  The \emph{bond-weighted} mean then drifts upward
with $w_0$ where fragmentation is weak ($\langle\ell\rangle_B=8.8/18.3/33.1/47.5$ at
$(\alpha,V)=(2,t_0)$ for $w_0=10/30/100/300$; $5.6/7.2/7.9/7.5$ at $(2,2t_0)$), but this is a
tail effect and does not enter the object density in the pair count.  The \emph{count}-weighted mean and the
active-object density $A_{\rm isl}=\rho_{\rm act}/p$ (active clusters per resonant bond) both
saturate: $\langle\ell\rangle_{\rm count}=4.6/4.9/4.8/4.6$ and
$A_{\rm isl}=0.169/0.145/0.142/0.150$ at $(2,t_0)$; $4.3/4.5/4.3/4.2$ and
$0.184/0.161/0.162/0.165$ at $(1.5,t_0)$; $2.5/2.4/2.4/2.2$ and $0.284/0.279/0.274/0.289$ at
$(2,2t_0)$; $2.1/2.2/2.1/2.0$ and $0.301/0.294/0.307/0.317$ at $(1.5,2t_0)$; cluster medians are
$1$--$2$ bonds throughout.  The object-density scaling is thus controlled and the integrated
transition weight per length is bounded by a sum rule; the low-frequency matching density of the
asymptotic rare tail is bounded by neither, shows no growth over the accessible cluster sizes,
and its uniform boundedness along the tail is an assumption of the construction~\cite{companion}.  Over the accessible range the tail is weakly coupled: the doublet transition norm
falls with cluster size, $\langle\lVert m\rVert\rangle=0.67,0.54,0.46,0.42,\ldots,0.28$ for
$\ell=1,2,3,4,\ldots,12$ at $(2,t_0)$: the rare large clusters are also the most weakly coupled.
Recursive healing therefore broadens the cluster-size distribution while leaving the density of
typical active objects --- the quantity in $\rho_{\rm act}=A_{\rm isl}p$ --- $w_0$-independent.

\emph{The surviving single-bond fraction.}  The reconstruction leaves a finite population of
isolated wing bonds at every $w_0$.  The fraction $f_1$ of \emph{active} objects that are single
bonds is $0.352,0.365,0.376,0.403$ for the raw partition and $0.320,0.427,0.461,0.502$ at the
recursive fixed point, at $(\alpha,V)=(2,t_0)$ for $w_0=10,30,100,300$; at $(1.5,2t_0)$ it is
$0.576$--$0.599$ and $0.536$--$0.592$ respectively.  There is no downward trend, as the scaled
wing argument of the Letter requires, and this is what keeps $b_{\rm all}=f_1^2b_1>0$.  The
factorized sector weights also survive the comb correlations: measured on the mesoscopic shells
used for the typicality scans, the wing--wing pair fraction is $0.1779$ against
$f_1f_1=0.1682$ at $\alpha=1.5$ and $0.1468$ against $0.1430$ at $\alpha=2$ ($V=t_0$) --- ratios
$1.058$ and $1.027$.

\section*{S2.\ Island diagonalization and the matching-law diagnostic [Fig.~2(b)]}

\subsection*{Protocol}

Each island harvested from the ensemble above (at $w_0=100$) is diagonalized exactly in its own
particle-number sector: an island of $\ell$ bonds is $\ell+1$ sites carrying the sampled particle
number $m$; the local Hamiltonian contains the hopping $-t_0$, the exact onsite energies $h_i$
plus the frozen exterior Hartree field of the sampled configuration, and the exact internal
couplings $U_{ij}$ of the island sites.  Islands with $m=0$ or $m=\ell+1$ are frozen (no internal
doublet) and are excluded --- they are roughly half of all islands.  Dense diagonalization is
used for $\ell\le10$ and sparse Lanczos (two lowest states) for $\ell=11$--$20$, Hilbert-space
dimensions up to $352\,716$; islands beyond $\ell=20$ carry $5.2\%/7.4\%$ of the resonant-bond
weight at $V=t_0$ ($\alpha=1.5/2$) and $\le0.1\%$ at $V=2t_0$, and are excluded.  Each island
yields its splitting $\Omega=E_1-E_0$ and transition-density norm
$\lVert m\rVert=[\sum_a|\langle E_0|n_a|E_1\rangle|^2]^{1/2}$.

The splitting distribution of genuine islands is regular: for $\ell\ge3$, the median of $\Omega$
is $1.02$--$1.13\,t_0$, a fraction $0.37$--$0.49$ lies below $t_0$, and over $98\%$ lies below
$2t_0$ --- the two-level edge at $2t_0$ is absent.  For $\ell=1$ objects the edge is exact
($\Omega_{\min}=2t_0$), but the conditional detuning of the \emph{surviving} isolated bonds is
reshaped by fragmentation (they live preferentially in the run wings, biased toward
$|\Delta|\to t_0$), which reduces their matching-log slope below the ideal benchmark of
Sec.~S5.

\subsection*{Saturation of $D(w)$ and the residual single-bond logarithm}

The pair-density diagnostic $D(w)=\Pr(|\Omega_A-\Omega_B|<w)/2w$ of Fig.~2(b) uses the island
splittings alone; matching probabilities are evaluated over $2\times10^7$ independently sampled
pairs.  Fitting $P^{\rm match}/q=a+b\ln(1/q)$ over $q\le10^{-2}$ gives the slopes of
Table~\ref{smtab:slopes}: for genuine islands ($\ell\ge3$) the slope vanishes,
$b=0.00\pm0.02$ in all cases, while the full mixed ensemble retains a small slope carried
entirely by the surviving single-bond subpopulation,
$b_{\rm all}\simeq f_{\ell=1}^2\,b_{\ell=1}$; that subpopulation is not dilute
($f_{\ell=1}\simeq0.4$--$0.6$), which is why the ensemble logarithm survives.  The $D(w)$ saturation was additionally verified
for the $\ell=11$--$20$ sector separately (flat at $D\simeq1.7$--$2.1$, larger than the
small-island value but bounded --- large islands match more easily, but boundedly) and, at
$(\alpha,V)=(1.5,2t_0)$, by exact all-pairs counting over $8868$ islands:
$D=0.739,0.743,0.737,0.738,0.738,0.739,0.735$ at
$w/t_0=10^{-4},3{\times}10^{-4},10^{-3},3{\times}10^{-3},10^{-2},3{\times}10^{-2},10^{-1}$, with
statistical errors below $1\%$; $D$ is quoted in units of $t_0^{-1}$ throughout.

Two diagnostics close the island ensemble.  \emph{(i) The $\ell=2$ sector:} two-bond islands are
$f_{\ell=2}=0.234$ and $0.245$ of the population at $(\alpha,V)=(2,t_0)$ and $(1.5,2t_0)$, with
matching slopes $b_{\ell=2}=-0.008$ and $-0.005$ (over $10^6$ pairs each) --- zero within noise
--- and $b_{\ell\ge2}=-0.001$ for the combined genuine-island class: the logarithm is carried
exclusively by $\ell=1$.  \emph{(ii) All transitions:} because the parent problem is a
high-energy one, the diagnostic was repeated using each island's full many-body spectrum (dense
spectra for every $\ell\le6$ island plus a $20\%$ subsample of $\ell=7$--$10$; $10\,852$ and
$14\,546$ islands), retaining every eigenstate pair with local transition density
$W_{rs}=\sum_a|\langle r|n_a|s\rangle|^2>10^{-10}$ and weighting channels by $W_{rs}$ at uniform
(infinite-temperature) initial-state occupation.  Over $6\times10^6$ cross-island transition
pairs, $D$ remains flat over three decades --- $0.270$--$0.274$ with small-$w$ slope $+0.0001$
at $(2,t_0)$, and $0.283$--$0.292$ at $(1.5,2t_0)$: no singularity appears in the weighted
infinite-temperature all-transition ensemble, so the lowest doublet used in the main analysis is
representative, not special.

\begin{table}[h]
\caption{Matching-log slopes $b$ by pair class.  The full-ensemble slope obeys
$b_{\rm all}\simeq f_{\ell=1}^2\,b_{\ell=1}$, identifying the residual logarithm as the
contribution of the surviving single-bond fraction $f_{\ell=1}$.}
\label{smtab:slopes}
\begin{ruledtabular}
\begin{tabular}{ccccccc}
$\alpha$ & $V/t_0$ & $f_{\ell=1}$ & $b_{\ell=1}$ & $b_{\ell\ge3}$ & $b_{\rm all}$ & $f_{\ell=1}^2b_{\ell=1}$\\
\colrule
$1.5$ & $1$ & $0.44$ & $0.60$ & $-0.02$ & $0.11$ & $0.11$\\
$2$   & $1$ & $0.41$ & $0.28$ & $0.00$  & $0.048$ & $0.047$\\
$1.5$ & $2$ & $0.56$ & $0.75$ & $0.01$  & $0.24$ & $0.24$\\
$2$   & $2$ & $0.52$ & $0.74$ & $0.00$  & $0.20$ & $0.20$\\
\end{tabular}
\end{ruledtabular}
\end{table}

\subsection*{Matching law on the recursively healed ensemble}

The diagonalization above is performed on the snapshot island partition.  To verify that the
sector structure survives the recursive boundary healing of Sec.~S1, the healed
fixed-point clusters themselves were diagonalized for two representative parameter sets at
$w_0=30$ and $100$, with the same protocol (dense diagonalization, here for $\ell\le14$).  The
healed pool differs from the snapshot pool in exactly the expected way --- the bond-weighted tail
is broader, so a larger fraction of resonant-bond weight sits in clusters beyond the
diagonalization cap ($0.62$--$0.66$ at $(2,t_0)$ versus $0.07$ for $\ell>20$ in the snapshot;
$0.12$--$0.13$ at $(1.5,2t_0)$) --- but the matching statistics are unchanged in structure
(Table~\ref{smtab:healmatch}): the $\ell\ge2$ sector is slope-free, $D(w)$ shows no systematic low-frequency growth
over three decades, and the full-ensemble slope obeys $b_{\rm all}=f_1^2b_{\ell=1}$ to within $3\%$ with
$f_1$ the healed single-bond fraction, which at $w_0=100$, $(2,t_0)$ is $0.47$ against the
$0.461$ of Sec.~S1.  The $b_{\ell=1}$ values are somewhat below the snapshot ones
(the surviving singletons after healing sit still closer to the window edge), so the residual
logarithm is if anything weaker, but it is present at the fixed point with the same sector
origin.  The excluded large clusters are the rare tail whose low-frequency matching density is
the stated assumption of the construction; within $\ell\le14$ it shows no growth.

\begin{table*}[t]
\caption{Matching law recomputed on the recursively healed fixed-point ensemble.  Healed
clusters of $\ell\le14$ bonds are diagonalized ($f_1$ is the single-bond fraction of all active
healed clusters; the fraction of resonant-bond weight in excluded clusters $\ell>14$ is listed);
$D(w)$ is the pair-density diagnostic of the $\ell\ge3$ sector at $w=10^{-4},10^{-3},10^{-2}$ in
units of $t_0^{-1}$; slopes fitted over $q\le10^{-2}$; $48$ chains of $4\times10^4$ bonds per row.}
\label{smtab:healmatch}
\begin{ruledtabular}
\begin{tabular}{ccccccccc}
$(\alpha,V/t_0)$ & $w_0$ & excl. & $f_1$ & $b_{\ell=1}$ & $b_{\ell\ge2}$ & $b_{\rm all}$ & $f_1^2b_{\ell=1}$ & $D(w)$\\
\colrule
$(2,1)$   & $30$  & $0.66$ & $0.44$ & $0.23$ & $-0.007$ & $0.069$ & $0.071$ & $0.67,0.83,0.79$\\
$(2,1)$   & $100$ & $0.62$ & $0.47$ & $0.19$ & $-0.001$ & $0.055$ & $0.055$ & $0.87,0.90,0.88$\\
$(1.5,2)$ & $30$  & $0.12$ & $0.53$ & $0.92$ & $-0.002$ & $0.264$ & $0.266$ & $0.62,0.62,0.62$\\
$(1.5,2)$ & $100$ & $0.13$ & $0.54$ & $0.78$ & $0.002$  & $0.237$ & $0.235$ & $0.66,0.62,0.62$\\
\end{tabular}
\end{ruledtabular}
\end{table*}

\section*{S3.\ Island-level shell pair-count distributions [Table~I]}

The typicality statistics of the Letter are evaluated at the level of the physical objects.  On
chains of $L=8000$ sites at $h/t_0=150$ (so the all-resonant onset scale
$i_*\simeq8.5\times10^4\gg L$), with the full detuning at $V=t_0$, every maximal resonant run in
the region of interest is identified, frozen runs are excluded, and each remaining island is
diagonalized exactly (dense to $\ell=10$, sparse to $\ell=14$; longer runs are rare at these
parameters and skipped).  Pairs are counted between islands centered in $x_0=4000\pm60$ and
islands in the one-sided shell $R=1800\pm2\ell_{\rm comb}$, with the matching condition
$|\Omega_A-\Omega_B|<2\lambda(V/R^\alpha)|g|\,\lVert m_A\rVert\lVert m_B\rVert$, $g$ a Gaussian
amplitude.  The scale factor $\lambda$ is an auxiliary rescaling parameter used to compare the
distributions at matched mean --- it absorbs the $\order(1)$ matching constants \emph{and}
compensates the fixed-shell $R^{-\alpha}$ factor between different $\alpha$, and is not
interpreted as a physical coupling prefactor --- tuned so that $\langle N_{\rm pair}\rangle$
sweeps through the takeoff value; $600$ realizations per $(\alpha,\lambda)$.  Table~\ref{smtab:typ} gives the scans at $\alpha=0.5$ (isolated-bond
regime), $\alpha=1$ (marginal point), and $\alpha=1.5$ (fragmented regime); the matched-mean rows
$\langle N\rangle\simeq1.4$ are those of Table~I of the Letter.

Down to $\langle N\rangle\simeq0.2$ the probability of at least one matched pair tracks the
Poisson reference to within $\simeq15\%$, and the variance exceeds it by less than a factor of
two, in all three regimes: the mean is not carried by rare comb coincidences, and the three
distributions are closely comparable at matched mean.

The $\alpha$ labels of the main-text table are the asymptotic classification; at the parameters
above the bare runs have $w_0(x_0)\simeq14$.  Repeating the $\alpha=0.5$ scan in the
one--two-bond regime --- $h=1100\,t_0$, so $w_0(x_0)\simeq1.9$ and the harvested objects are
one--two-bond ($2400$ realizations, center window $\pm150$) --- gives the same structure at
matched mean: $\langle N\rangle=1.41$, median $1$, $P(N\ge1)=0.61$ $[0.76]$,
${\rm Var}/\langle N\rangle^2=1.44$ $[0.71]$, and the same monotone $\lambda$ scan
($\langle N\rangle=0.18/0.33/0.58/1.00/1.67$ at $\lambda=0.3/0.6/1.2/2.4/4.8$, medians
$0/0/0/1/1$).  Typicality is not an artifact of the multi-bond parameters.

The Gaussian amplitude is a surrogate; repeating the matched-mean rows with the matching window
built from the \emph{physical} conditional island coupling,
$|\Omega_A-\Omega_B|<2\lambda(V/R^\alpha)|\Gamma_{AB}|$ with
$\Gamma_{AB}=\sum_{ab}X_{ab}\,m^A_am^B_b$ and $X_{ab}$ independent uniform on $[-1,1]$ (leading
order in $R$), gives closely comparable distributions (Table~\ref{smtab:typG}): the typicality
conclusion does not depend on the surrogate amplitude.

\begin{table}[h]
\caption{Matched-mean shell pair-count distributions with the physical island form-factor
coupling $\Gamma_{AB}$ replacing the Gaussian amplitude ($V=t_0$, $600$ realizations per row).
Reference values as in Table~\ref{smtab:typ}.}
\label{smtab:typG}
\begin{ruledtabular}
\begin{tabular}{cccccc}
$\alpha$ & $\langle N\rangle$ & median & $P(N{\ge}1)$ & $P^{\rm Poi}$ & ${\rm Var}/\langle N\rangle^2$\\
\colrule
$0.5$ & $1.43$ & $1$ & $0.66$ & $0.76$ & $1.11$\\
$1.0$ & $1.39$ & $1$ & $0.64$ & $0.75$ & $1.41$\\
$1.5$ & $1.43$ & $1$ & $0.65$ & $0.76$ & $1.16$\\
\end{tabular}
\end{ruledtabular}
\end{table}

\begin{table}[h]
\caption{Island-level shell pair-count distributions ($V=t_0$, $600$ realizations per row).
$P^{\rm Poi}$ and $r^{\rm Poi}$ denote the Poisson reference values $1-\e^{-\langle N\rangle}$
and $1/\langle N\rangle$.}
\label{smtab:typ}
\begin{ruledtabular}
\begin{tabular}{cccccccc}
$\alpha$ & $\lambda$ & $\langle N\rangle$ & med. & $P(N{\ge}1)$ & $P^{\rm Poi}$ & ${\rm Var}/\langle N\rangle^2$ & $r^{\rm Poi}$\\
\colrule
$0.5$ & $0.05$ & $1.44$ & $1$ & $0.69$ & $0.76$ & $1.13$ & $0.70$\\
$0.5$ & $0.1$  & $2.60$ & $2$ & $0.86$ & $0.93$ & $0.79$ & $0.38$\\
$0.5$ & $0.2$  & $4.67$ & $4$ & $0.94$ & $0.99$ & $0.55$ & $0.21$\\
$0.5$ & $0.4$  & $8.30$ & $7$ & $0.99$ & $1.00$ & $0.39$ & $0.12$\\
\colrule
$1$ & $4$  & $0.42$ & $0$ & $0.30$ & $0.34$ & $3.4$ & $2.4$\\
$1$ & $8$  & $0.78$ & $0$ & $0.46$ & $0.54$ & $2.0$ & $1.3$\\
$1$ & $16$ & $1.41$ & $1$ & $0.64$ & $0.76$ & $1.24$ & $0.71$\\
$1$ & $32$ & $2.52$ & $2$ & $0.80$ & $0.92$ & $0.83$ & $0.40$\\
$1$ & $48$ & $3.56$ & $3$ & $0.87$ & $0.97$ & $0.66$ & $0.28$\\
\colrule
$1.5$ & $320$  & $0.31$ & $0$ & $0.26$ & $0.27$ & $3.4$ & $3.2$\\
$1.5$ & $640$  & $0.59$ & $0$ & $0.40$ & $0.45$ & $2.1$ & $1.7$\\
$1.5$ & $1280$ & $1.12$ & $1$ & $0.57$ & $0.67$ & $1.42$ & $0.89$\\
$1.5$ & $1600$ & $1.38$ & $1$ & $0.63$ & $0.75$ & $1.23$ & $0.73$\\
$1.5$ & $2560$ & $2.08$ & $2$ & $0.74$ & $0.88$ & $0.99$ & $0.48$\\
\end{tabular}
\end{ruledtabular}
\end{table}

\section*{S4.\ Exact resonant Fock graph at $L\le18$ [End Matter]}

For each realization $(\phi,\{V_{ij}\})$, all $\binom{L}{L/2}$ half-filled configurations are
enumerated, the full detuning $\Delta_a(S)$ is evaluated for every bond of every configuration,
and every active edge (flippable and $|\Delta_a|<t_0$) is inserted; connected components are
found exactly, and the forward branching $B$ is evaluated from the identical samples.
Realization counts are $300/200/100/50$ for $L=12/14/16/18$; $V=t_0$; the field grid is
$h=f\,h_0(L)$ with $h_0(L)$ set by the supply law, $f=0.55$--$1.5$.

The $B=1$ crossings are at $N_{\rm res}=2.41/2.41/2.42/2.42$ ($\alpha=0.5$, $L=12/14/16/18$),
$2.51/2.54/2.52/2.55$ ($\alpha=1$), and $2.60/2.61/2.62/2.65$ ($\alpha=2$).  At these crossings
the largest component holds $13$--$53$ vertices and the mean finite-cluster size is
$\chi_{\rm cl}=3.3$--$5.0$, essentially independent of $L$.  Table~\ref{smtab:fock} shows the
percolation diagnostic $S_{\rm max}/N_F^{2/3}$ ($N_F$ the number of Fock-space vertices;
$N_F^{2/3}$ the mean-field critical-window reference scale) at $\alpha=0.5$ and $2$.  At every
fixed $N_{\rm res}$ the diagnostic falls with $L$: the finite-size graphs are far from any
giant-component regime, and the data are incompatible with ordinary percolation at fixed
$\order(1)$ resonance count; extrapolating to strict thermodynamic subcriticality would exceed
what $L\le18$ can establish.

\begin{table}[h]
\caption{Percolation diagnostic $S_{\rm max}/N_F^{2/3}$ of the exact resonant Fock graph
($V=t_0$).  Rows are the field grid $f=h/h_0(L)$; $N_{\rm res}$ varies weakly with $L$ at fixed
$f$ (values quoted for $L=18$).}
\label{smtab:fock}
\begin{ruledtabular}
\begin{tabular}{ccccccc}
$\alpha$ & $f$ & $N_{\rm res}$ & $L{=}12$ & $14$ & $16$ & $18$\\
\colrule
$0.5$ & $0.55$ & $4.75$ & $2.63$ & $2.34$ & $2.20$ & $1.52$\\
$0.5$ & $0.70$ & $3.64$ & $1.15$ & $0.77$ & $0.56$ & $0.27$\\
$0.5$ & $0.85$ & $2.87$ & $0.58$ & $0.33$ & $0.20$ & $0.09$\\
$0.5$ & $1.00$ & $2.36$ & $0.34$ & $0.17$ & $0.09$ & $0.04$\\
$2$   & $0.55$ & $4.69$ & $0.92$ & $0.43$ & $0.17$ & $0.07$\\
$2$   & $0.70$ & $3.48$ & $0.36$ & $0.17$ & $0.06$ & $0.03$\\
$2$   & $0.85$ & $2.81$ & $0.19$ & $0.09$ & $0.03$ & $0.01$\\
$2$   & $1.00$ & $2.39$ & $0.12$ & $0.06$ & $0.02$ & $0.01$\\
\end{tabular}
\end{ruledtabular}
\end{table}

\section*{S5.\ The isolated-bond matching benchmark}

For one resonant bond the conditional detuning is uniform to leading order in the
resonance-starved regime, which implies the normalized splitting density
\begin{equation}
 f(\Omega)=\frac{\Omega}{t_0\sqrt{\Omega^2-4t_0^2}},
 \qquad 2t_0<\Omega<\sqrt5\,t_0.
\label{eq:fOmega}
\end{equation}
The square-root lower-edge singularity is the Jacobian of the map $\delta\mapsto\Omega$ and
survives smooth deformations of the single-bond detuning distribution.  If two resonant bonds
are drawn independently from Eq.~\eqref{eq:fOmega}, the physical mixing-angle resonance
condition $|\Omega_1-\Omega_2|<|J^{zz}|(2t_0/\Omega_1)(2t_0/\Omega_2)$ gives, for
$q=|J^{zz}|/t_0\ll1$,
\begin{align}
 P_{\rm mix}^{\rm(ind)}(q)&=2q\Big[\ln\frac1q+c_{\rm mix}\Big]+o(q),\nonumber\\
 c_{\rm mix}&=1+\ln\!\left[\frac{16(5\sqrt5-8)}{28+13\sqrt5}\right]=0.8853\ldots
\label{eq:Pmixind}
\end{align}
The convention matters: widening the window by a factor $\eta$ changes both the leading
coefficient and the constant, so the slope $2$ is a benchmark constant of the isolated-bond
regime for the window convention $\eta=1$, not a universal physical number.  Averaging over the
random four-term coupling amplitude shifts only the finite constant,
$c_{\rm mix}\to c_\star=c_{\rm mix}-\tfrac{15}{14}(\ln4-\tfrac{539}{450})=0.68334\ldots$, using
the exact moments $\mathbb E|\varsigma|=14/15$ and
$\mathbb E[|\varsigma|\ln|\varsigma|]=\ln4-539/450$ of the sum of four independent $U[-1,1]$
variables~\cite{companion}.

Two direct checks fix the interpretation of the logarithm.  For a \emph{fixed} physical pair at
$V=0$, averaging over the global phase alone, the conditional matching probability is strictly
linear ($P/q=4.795+0.0033\ln(1/q)$ over three decades): one pair is a one-dimensional constraint
and carries no logarithm.  Averaging over the pair \emph{ensemble} ($L=100$, $h/t_0=25$,
$\alpha=1$, all separations $R\ge2$) reproduces Eq.~\eqref{eq:Pmixind} with no free parameter:
fitted slopes $2.00(5)$ at $V=0$ and $1.95(4)$ at $V=t_0$ against the predicted $2$.  The
logarithm is a property of the bond ensemble sharing the deterministic band edge $2t_0$ --- and
it is precisely this shared edge that fragmentation removes \emph{from the multi-site objects}
(Sec.~S2), replacing their intrinsic law by the linear island law.  The surviving wing bonds
retain the edge, so the full ensemble keeps a reduced logarithm, $b_{\rm all}=f_1^2b_1$.

\section*{S6.\ Direct fixed-shell test of the pair-count scaling}

The scaling law of the Letter,
$\Lambda(x)\sim(Vt_0/h^2)\,x^{4-2n-\alpha}[B\ln x+C]$ [Eq.~(9)], is tested directly, with the
algebraic exponent fixed in advance.  On chains with the full detuning, active islands are
harvested in a centre window of $121$ sites at $x_0$ and in a one-sided shell at
$R=\kappa x_0$, $\kappa=0.45$, of width $4\ell_{\rm comb}$; each active island enters with its
exact doublet $(\Omega,\lVert m\rVert)$, and pairs are matched with the physical form-factor
coupling, $|\Omega_A-\Omega_B|<2\lambda(V/R^\alpha)|\Gamma_{AB}|$.  The matching scale $\lambda$
is fixed \emph{per $\alpha$} (not per $x$ or $h$) such that the window is dilute,
$q\le0.02$ at the smallest shell; the measured $\Lambda=n_RR$ then gives
$Y\equiv\Lambda h^2x_0^{-(4-2n-\alpha)}/(Vt_0)$, which the theory predicts to be
$h$-independent and linear in $\ln x$ at fixed $(\alpha,V)$, with nonuniversal
$B_{\alpha,V}>0$ and $C_{\alpha,V}$.  Parameters: $x_0=4000$--$24\,000$,
$h/t_0=100$--$340$ ($3000$ realizations per point), $V=t_0$, $\alpha=0.5$, $1$, $1.5$ ---
the bond side, the marginal point, and the fragmented side.

\begin{figure}[t]\centering
\includegraphics{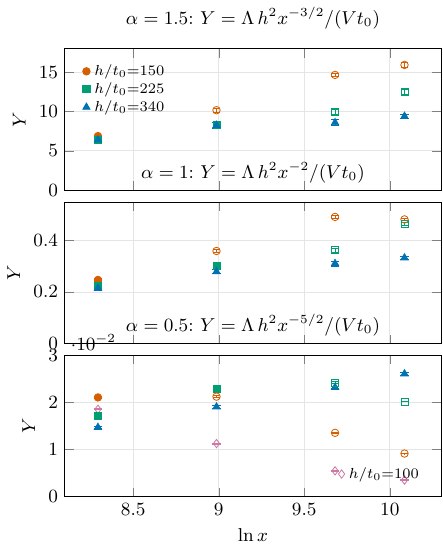}
\caption{Direct fixed-shell test of the pair-count scaling, Eq.~(9) of the Letter:
$Y\equiv\Lambda\,h^2x^{-(4-2n-\alpha)}/(Vt_0)$ against $\ln x$ at fixed shell fraction
$\kappa=R/x=0.45$ and fixed matching scale per $\alpha$ (dilute window, $q\le0.02$ at the
smallest shell), with the algebraic exponent \emph{fixed in advance}, not fitted.  Filled
symbols: points inside the resonance-starved window $w_0/\ell_{\rm comb}\le0.15$; open symbols:
outside it.  Batch-mean error bars (10 batches; often smaller than the symbols).}
\label{smfig:lamtest}\end{figure}

Two domain conditions bound the test and were fixed before the analysis: diluteness of the
matching window, and the resonance-starved condition $w_0\ll\ell_{\rm comb}$, implemented as
$w_0/\ell_{\rm comb}\le0.15$.  Figure~\ref{smfig:lamtest} shows the result.  Within the starved
window the $h^{-2}$ rescaling collapses the series to $7\%$/$1\%$ at $\alpha=1.5$,
$14\%$/$8\%$ at $\alpha=1$, and $33\%$/$18\%$ at $\alpha=0.5$; $Y$ grows with $\ln x$ with
positive fitted slope at every $\alpha$ ($B=1.53(12)$, $0.059(4)$, $0.0037(1)$); and the free
secondary power-law fit gives slopes $1.65$, $2.15$, $2.74$ against the prescribed $1.5$, $2$,
$2.5$ --- an excess consistent in sign and scale with the logarithmic correction plus
finite-window effects, though the accessible $\ln x$ interval ($\simeq1.8$ inside the window)
does not distinguish $B\ln x+C$ from a more general slowly varying correction.  Most
importantly, the fragmented side collapses at least as well as the bond-side control: there is
no deterioration that grows with $w_0$ at $\alpha=1.5$, which is the failure mode a breakdown
of the island construction by the large-cluster tail would produce.

\begin{figure}[t]\centering
\includegraphics{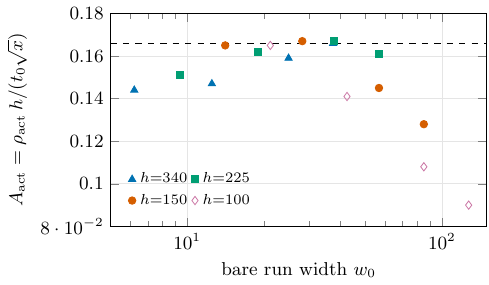}
\caption{Normalized active-object density $A_{\rm act}$ at $\alpha=0.5$, $V=t_0$, against $w_0$.
At small $w_0$, $A_{\rm act}$ rises toward a common plateau $\simeq0.166$ (dashed) reached for
$w_0\gtrsim15$; it then falls once the starvation ratio $w_0/\ell_{\rm comb}$ exceeds
$\simeq0.2$ (the falling points are those with $w_0/\ell_{\rm comb}=0.21$--$0.51$),
independently of how that ratio is reached.  The residual spread of
Fig.~\ref{smfig:lamtest} is carried by this factor squared; the matching fraction itself is
$h$-independent to $10$--$15\%$ throughout.}
\label{smfig:Aact}\end{figure}

Outside the starved window the collapse breaks in both directions --- $Y$ rises by up to
$60\%$ at $\alpha=1.5$ and falls by up to a factor $3$ at $\alpha=0.5$.  The origin was
localized by decomposition: the matching fraction at fixed window is $h$-independent to
$10$--$15\%$ at every $x$ (and its $R^{\alpha}$-scaled value grows with $\ln R$), while the
active-object density departs from its assumed $\rho_{\rm act}\propto t_0\sqrt{x}/h$ form.
Figure~\ref{smfig:Aact} isolates this factor at $\alpha=0.5$, including a dedicated $h=100$
series: at fixed $x$, decreasing $h$ while remaining within the resonance-starved window
increases $w_0$ and brings the normalized active-object density toward its asymptotic value
($A_{\rm act}\simeq0.144\to0.166$ over $w_0=6\to40$), identifying the residual spread as a
local-object finite-$w_0$ correction rather than a failure of the pair-matching scaling; past
$w_0/\ell_{\rm comb}\simeq0.2$ the density falls, marking the end of the starved regime.  The
breakdown is therefore a loss of the object-density asymptotics at the window boundaries, not
of the long-range matching mechanism.

\end{document}